 %% Plain.tex

\magnification=\magstep1
\baselineskip=15pt
\overfullrule=0pt

\def\N{{\cal N}}
\def\O{{\cal O}}

\def\half{{1 \over 2}}
\def\d{{d \over 2}}
\def\k{{k \over 2}}
\def\AdS{{\rm AdS}}

\def\square{{\sqcap \! \! \! \! \sqcup}}
\def\leftrightharpoon{\buildrel \leftrightarrow \over}
\font\fourteenbf=cmbx10 scaled \magstep2

\input BoxedEPS
%\SetTexturesEPSFSpecial  %% for the Mac & Textures
\SetRokickiEPSFSpecial  %% for dvips by Tom Rokicki, for VMS or UNIX?
\HideDisplacementBoxes

\rightline{UCLA/98/TEP/33}
\rightline{MIT-CTP-2781}

\bigskip

\centerline{{\fourteenbf GAUGE BOSON EXCHANGE IN AdS$_{{\bf
d+1}}$}\footnote{*}{Research supported in part by the National Science
Foundation under grants PHY-95-31023,  PHY-94-07194 and  PHY-97-22072.}}

\bigskip
\bigskip
\bigskip

\centerline{{\bf Eric D'Hoker}${}^1$ 
            {\bf and Daniel Z. Freedman}${}^2$}

\bigskip

\centerline{${}^1$Department of Physics}
\centerline{University of California, Los Angeles, CA 90024, USA;}

\bigskip

\centerline{${}^2$Department of Mathematics and Center for Theoretical Physics}
\centerline{Massachusetts Institute of Technology, Cambridge, MA 02139, USA}

\bigskip
\bigskip
\bigskip

\centerline{\bf ABSTRACT}

\bigskip
\noindent
 We study the amplitude for exchange of massless gauge bosons between
  pairs of massive scalar fields in Anti-de Sitter space.  In the
  AdS/CFT correspondence this amplitude describes the
  contribution of conserved flavor symmetry currents to 4-point
  functions of scalar operators in the boundary conformal theory.
  A concise, covariant, Y2K compatible derivation of the gauge boson
  propagator in $\AdS_{d+1}$ is given.  Techniques are developed
  to calculate the two bulk integrals over AdS space leading to
  explicit expressions or convenient, simple integral
  representations for the amplitude.  The amplitude contains leading power and 
  sub-leading logarithmic singularities in the gauge boson channel and
  leading logarithms in the crossed channel.  The new methods
  of this paper are expected to have other applications in the
  study of the Maldacena conjecture.

\vfill\break

\centerline{\bf I. Introduction}

\bigskip
\noindent
Many 2- and 3-point correlation functions have been calculated 
in studies of the Maldacena conjecture [1--3], and attention  has
recently been turned to 4-point correlators [4--8]. One of the goals is to obtain non-perturbative information
about the large $N$, fixed large $g^2 N$ limit of the $\N =4$
super--Yang-Mills theory with gauge group $SU (N)$.  One important question
is whether the theory has a simple $t$-channel OPE structure,
so that 4-point functions have convergent expansions of the
schematic form [9]
$$
\eqalign{
    \langle \O _1 (x_1) & \O _2 (x_2) \O_3 (x_3) \O_4 (x_4) \rangle  \cr
&   = \sum_p {\gamma_{1 3 p} \over (x_1-x_3)^{\Delta_1 + \Delta_3 - 
      \Delta_p}} ~
    {1 \over (x_1-x_2)^{2 \Delta _p}} ~
    {\gamma_{2 4 p} \over (x_2-x_4)^{\Delta_2 + \Delta_4 - \Delta_p}}
\cr}
\eqno (1.1)
$$
containing the contribution of a finite number of primary
operators (and descendents).  A related question is whether
4-point correlators in the large $N$ supergravity
 approximation are given by their free-field values as is the
 case for 3-point functions [10,11].  The
 discovery [7] of logarithmic singularities in some diagrams 
contributing to
 the correlator $\langle \O_\phi \O_c \O_\phi \O_c \rangle$ of
 operators corresponding to the bulk dilaton $\phi$ and
 axion $c$ fields suggests that the large $N$ limit is more
 complicated than the simple picture suggested by (1.1). 
 However, a definite answer is not yet known because neither $\langle 
 \O _\phi \O _c \O _\phi \O _c \rangle$  nor any other 4-point
 correlator has been completely calculated, and logarithms may cancel
when all diagrams are included in the full amplitude.

Four-point correlators depend continuously on two conformal
invariant variables.  Thus they are inherently more complex than
2- and 3-point functions whose form is determined up to a
small number (typically 1) of constants by conformal symmetry.
The study of 4-point correlators from the AdS/CFT
correspondence has been hampered by several difficulties.

\item{1.} Simple covariant expressions for the bulk-to-bulk gluon and 
  graviton propagators in $\AdS_{d+1}$ are not known.

\item{2.}  The integrals required to compute diagrams containing a
  bulk-to-bulk propagator (even for a scalar) are difficult, and 
  general techniques to evaluate them have not yet been
  developed.

\item{3.} The computation of realistic 4-point correlators in the
  $\N=4$ super--Yang-Mills theory requires the specific values of cubic and
  quartic coupling in the Type IIB, $d=10$ supergravity theory on
  $\AdS_5 \times S^5$.  Although the complete mass spectrum of this theory is    
known [12], there seems to be rather little information on the couplings. The  
cubic couplings of scalars corresponding to conformal primary operators are one 
happy exception [10].

In this paper we address the first two issues discussed above.  In
\S 2, we give a simple, covariant derivation of
the gauge boson propagator.  In \S 3, we consider the
AdS integral for the gauge boson exchange contribution (see
Fig.\thinspace1) to the 4-point function correlator $\langle \O_{\Delta} (x_1)
\O_{\Delta'} (x_2) \O^*_{\Delta} (x_3) \O^*_{\Delta'} (x_4)
\rangle$ of charged scalar operators of scale dimensions $\Delta$
and $\Delta'$.  The required cubic coupling is determined by
gauge symmetry. We develop expansion and resummation techniques
to evaluate the two integrals over $\AdS_{d+1}$ required to
calculate the diagram in Fig.\thinspace1.  This leads to a
rather simple 2-parameter integral representation of the
amplitude which we pursue toward explicit evaluation in
\S 4.  We establish that the correlator contains a
leading logarithmic $s$-channel singularity in the variable $x_{1
2} $ (we use the notation $x_{ij}=x_i-x_j$) or $x_{3 4}$, 
and the same in the
$u$-channel variables $x_{1 4} $ or $x_{2 3}$.  In the
$t$-channel we find the leading singularity $1/(x_{2 4})^{2
  \Delta' -d+1}$ expected for a conserved current in the boundary
theory, but we also find a non-leading logarithmic term 
$$
x_{2 4}\,  \ln x_{24} .
\eqno (1.2)
$$  
The latter
provides further evidence (see also [7]) that the claimed
exact correspondence [6] between an AdS exchange
diagram and the contribution of a single primary and its
descendents in the OPE (1.1) is not completely
correct.

Because of the problem of unknown couplings our new 
results for gauge boson exchange amplitude are still not enough to give 
a complete calculation of any 4-point correlators of the $\N
=4$ theory.  It is possible that the logarithmic singularities of 
the gauge boson diagram cancel with those in other diagrams.  Indeed
we expect that the techniques developed here can be applied to the 
calculation of the integrals in other diagrams,\footnote{*}{In a recent MIT
seminar, H.~Liu outlined a treatment of diagrams with scalar exchange based on
the Mellin-Barnes representation of the hypergeometric function. This method
appears complementary to ours.} and we also hope that the approach used to 
obtain
the gauge boson propagator can be extended to the more complicated case of the
bulk-to-bulk propagator of the graviton.
\midinsert
$$
\BoxedEPSF{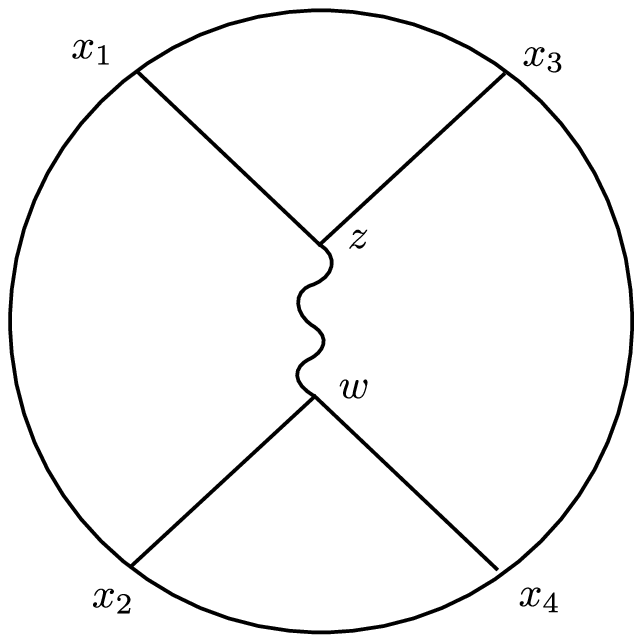 scaled 600}  %% scales it 66.6 percent
$$
\centerline{Figure\thinspace 1}
\endinsert

\bigbreak

\centerline{{\bf 2. The Gauge Boson Propagator}}

\medskip
\noindent
We work on the euclidean continuation of $\AdS_{d+1}$ which is
defined as the $Y_{-1}>0$ sheet of the hyperboloid
$$
  -Y^2_{-1} + \sum ^d_{i=0} Y^2_i =-r^2_0 
\eqno (2.1)
$$
embedded in a $d+1$ dimensional space with metric of
signature $(-++ \cdots +)$.  The AdS scale $r_0$ is set to $r_0=1$
in the following.  Defining $1/z_0 \equiv Y_{-1} + Y_0$ 
  and $z_i \equiv z_0 Y_i \quad i=1, \ldots ,d$, gives a complete
coordinate chart $z_{\mu} (Y)$ such that the induced metric on the hyperboloid
takes the form (with $ z_0 >0$)
$$  
  ds^2 = \sum _{\mu, \nu=0} ^ d g_{\mu \nu} dz_\mu dz_\nu
       ={1 \over z_0^2} (dz^2_0 + \sum ^d_{i=1} \, dz^2_i)
\, .
\eqno (2.2)
$$
This is a constant curvature metric with Ricci tensor $R_{\mu \nu}
=-d g_{\mu \nu}$.  It is well known that scalar propagators are most 
simply expressed [13] as (hypergeometric) functions of the 
chordal distance  $u$ between observation and source
points on the hyperboloid.  This can be written in terms of the
coordinates $z_{\mu}(Y)$ and $w_{\mu}(Y')$ as
$$
  u \equiv \half (Y-Y')^2 =
  {(z-w)^2 \over 2z_0 w_0}
\eqno (2.3)
$$
where $(z-w)^2 = \delta_{\mu \nu}(z-w)_{\mu} (z-w)_{\nu}$.

\medskip

The previous literature on the gauge field propagator in constant 
curvature space-times concentrates on the positive curvature de
Sitter space, although there is work, both old [14] and
new [6], on AdS.  In the first of these, the covariant
propagator is obtained in Feynman gauge and involves
transcendental functions.\footnote{*}{E.g., $\log u$ for odd $d$  and $\arcsin
\sqrt u$ for even $d$.}  These do
not appear for the physical components of the field in our
approach, which suggests that the method of [14] is too
entwined with gauge artifacts.  In [6] the Coulomb gauge
is used to find a quite simple expression for the propagator.
However this approach leads to an expression for the photon
exchange amplitude which is not manifestly covariant under the
isometries of $\AdS_{d+1}$ (and appears to be non-local as well).
Having criticized these earlier treatments, it must also be said
that we build upon them.  The emphasis of [14] on a
simple basis of independent bitensors is important for us as is 
the observation of [6] that the transverse, spatial modes
of the field have a simple scalar propagator.

\medskip

Let us begin with the action of an abelian gauge field coupled to 
a conserved current source in the $\AdS_{d+1}$ background:
$$
  S_A = \int d^{d+1}z \sqrt g \big [ {1 \over 4} F^{\mu \nu} F_{\mu \nu} +
      {\xi \over 2} (D_{\mu}A^{\mu})^2 +
      A_{\mu} J^{\mu} \big ] \, .
\eqno (2.4)
$$
In the gauge fixing term of $S_A$, $D_\mu$ is the AdS-covariant derivative.
The propagator is a bitensor $G_{\mu \nu'} (z,w)$ which satisfies
the AdS-covariant equation $(\partial_{\mu} =
{\partial \over \partial z^{\mu}})$
$$
  D^{\mu} \partial_{[\mu} G_{\nu ] \nu'} + \xi \partial_{\nu}
  (D^{\mu}G_{\mu \nu'}) =
  {g_{\nu \nu'} \over \sqrt{g}} \delta^{(d+1)}(z,w) \, .
\eqno (2.5)
$$
Any bitensor can be expressed [14] as a sum of two linearly 
independent forms with scalar coefficients.  We shall choose as
independent bitensors, two forms which are closely related to the 
biscalar variable $u$, namely (with $\partial _{\nu'} =
{\partial \over \partial w^{\nu'}}$)
$$
  \partial_{\mu} \partial_{\nu'}u= -
  {1 \over z_0w_0} \big [ \delta_{\mu \nu'} +
  {1 \over w_0} (z-w)_{\mu} \delta_{\nu'0} +
  {1 \over z_0} (w-z)_{\nu'}\delta_{\mu 0} -
  u \delta_{\mu 0} \delta_{\nu'0} ]
\eqno (2.6)
$$
and $\partial_{\mu}u \partial_{\nu'}u$ with
$$
\eqalign{
  \partial_{\mu} u &= {1 \over z_0}
     [(z-w)_{\mu} / w_0 - u \delta_{\mu 0}] \cr
   \partial_{\nu'} u &= {1 \over w_0}
     [(w-z)_{\nu'} / z_0 - u \delta_{\nu' 0}] \, .\cr}
\eqno (2.7)
$$

One innovation of our approach is the Ansatz for the propagator
$$
  G_{\mu \nu'}(z,w) =- (\partial_{\mu} \partial_{\nu'}u)
  F(u) + \partial_{\mu} \partial_{\nu'} S(u)
\eqno (2.8)
$$
where $F(u)$ and $S(u)$ are unknown scalar functions.  This ansatz
contains the two independent bitensors above.  However $S(u)$ is
clearly a gauge artifact which gives vanishing contribution when 
integrated against conserved currents in a Witten diagram.  (We must
also check that the surface term in the partial integration
vanishes.)  Thus we concentrate on the determination of $F(u)$,
which turns out to be quite simple, and have only secondary
concern for $S(u)$, just enough to be sure that the method is
consistent.

\medskip

To find $F(u)$, we write the equation of motion of (2.4) 
in full detail as
$$
\eqalign{
 \partial_{\mu} [z_0^{-d+3}(\partial_{\mu} A_{\nu}  - & \partial_{\nu}
  A_{\mu})] + \xi \partial_{\nu} [z_0^{-d+3} \partial_{\mu}A_{\mu}
  - (d-1) z^{-d+2}_0 A_{0}]  \cr
&       + (d-1) \xi \delta_{\nu 0} [z^{-d+2}_0 \partial_{\mu} A_{\mu}
       - (d-1) z^{-d+1}_0 A_0] = z^{-d+1}_0 J_{\nu} \, .
\cr}
\eqno (2.9)
$$
Consider the transverse spatial components
$$
  A^{\perp}_i \equiv (\delta_{ij} - \partial_i
  {1 \over \nabla^2} \partial_j) A_j 
\eqno (2.10)
$$
for which we use the temporary non-covariant notation $i \equiv
1, \ldots ,d$ with flat $\nabla^2$.  They satisfy the
gauge-independent equation
$$
( \partial^2_0 + \nabla^2) A^{\perp}_i - (d-3) {1 \over z_0}
     \partial_0 A^{\perp}_i = {1 \over z^2_0}J^{\perp}_i \, .
\eqno (2.11)
$$
As observed in the Coulomb gauge treatment of [6], this
means that $z_0 A^{\perp}_i (z)$ satisfies the same equation as a
scalar field of mass $m^2=-(d-1)$.  The solution
of (2.11) can then be written as
$$
  z_0 A^{\perp}_i (z) = \int {d^{d+1}w \over w^{d+1}_0}
  G(u(z,w)) w_0 J^{\perp}_i (w)
\eqno (2.12)
$$
where $G(u)$ is the scalar Green's function, which satisfies
$$
  D^{\mu} \partial_{\mu} G+ (d-1) G
  = {1 \over \sqrt{g}} \delta^{(d+1)} (z,w) \, .
\eqno (2.13)
$$
Let us compare (2.12) with the covariant solution
of (2.9), namely
$$
A_{\mu}(z) = \int {d^{d+1}w \over w^{d+1}_0}
G_{\mu \nu'} (z,w) J^{\nu'} (w) \, .
\eqno (2.14)
$$
We insert the representation (2.8), and observe that the
$\partial_{\mu} \partial_{\nu'}S$ term vanishes since the current
is conserved.  We apply the transverse spatial
projector (2.10) to both sides of (2.14), note that 
$$
{\partial \over \partial z_i}F(u) 
=-{\partial \over \partial w_i}F(u)
$$ 
as a consequence of the translation invariance of (2.9), and use
$$
  {(z-w)_i \over w_0} F(u) = {\partial \over \partial z_i}
  \int^u \, dv F(v) \, .
\eqno (2.15)
$$
The result is an equation for $z_0 A^{\perp}_i (z)$ which is
compatible with (2.12) provided we identify $F(u) \equiv 
G(u)$.  Thus the physical part of the covariant propagator is,
for any value of $\xi$, the scalar propagator of (2.13).

\medskip

To find the explicit solution of (2.13), one can refer
to the older AdS literature [13] (see also [4]), but we shall
proceed here in a self-contained way because we will need
similar techniques for other purposes below.  Typically one needs
to use hypergeometric transformation formulae to convert the results in the
literature to the simple explicit form we will find.

\medskip

We wish to express (2.13) as a differential equation in
the variable $u$.  For this we need certain properties of
derivatives of $u$, which we now simply list:
$$
\eqalign{
     \square u= D^{\mu} \partial_{\mu} u &= (d+1)(1+u) 
\cr
      g^{\mu \nu} \partial_{\mu} u \partial_{\nu}u &= u (2+u) 
\cr
      D_{\mu} \partial_{\nu} u &= g_{\mu \nu}(1+u) 
\cr    
      (D^{\mu}u)(D_{\mu} \partial_{\nu}\partial_{\nu'}u)
          &= \partial_{\nu}u \partial_{\nu'}u \, .\cr}
\eqno (2.16)
$$
These properties, some of which are not required immediately, can 
be derived with sufficient faith, perseverance, and Christoffel
symbols.  Using (2.16) one finds that
$$
\eqalign{
  D^{\mu} \partial_{\mu}F(u) 
&= g^{\mu \nu} \partial_{\mu} u \partial_{\nu} u F'' (u) 
    + \square u F' (u) \cr
&= u(2+u) F'' (u) + (d+1) (1+u) F' (u) \, .\cr}
\eqno (2.17)
$$
One can now check that the solution of (2.13) with the
fastest decay on the AdS boundary $(u \to \infty)$ is
$$
  F(u) = {\Gamma ({d-1 \over 2}) \over 4 \pi^{(d+1)/2}} \, 
         {1 \over [u (u+2)]^{(d-1)/2}}
\eqno (2.18)
$$
where the normalization is determined [13] by matching
to the short distance behavior in flat $(d+1)$-dimensional space.

\medskip

The overall consistency of this approach remains to be checked,
and that is done by substituting the ansatz (2.8)
into (2.5).  It is best to do this in two stages, so we
first try to find a choice of gauge parameter $\xi$ consistent
with $S(u) \equiv 0$.  Substitution of the $F(u)$ term
of (2.8) into (2.5) leads to the equation (for
separated points $z_{\mu} \neq w_{\mu}$)
$$
\eqalign{
     - (\partial_{\nu} \partial_{\nu'} u) D^{\mu} \partial_{\mu}F 
       + (\partial_{\mu} \partial_{\nu'}u) D^{\mu}\partial_{\nu}F
       - (D_{\mu} \partial_{\nu} \partial_{\nu'}u) D^{\mu}F \qquad\qquad&  
\cr
   + (D^{\mu} \partial_{\nu} \partial_{\nu'}u)F
      - \xi [ (\partial_{\nu} D^{\mu} \partial_{\mu} 
        \partial_{\nu'}u)F  
        + (D^{\mu} \partial_{\mu} \partial_{\nu'}u)
        \partial_{\nu}F {}\qquad&
\cr
   + (D_{\nu} \partial_{\mu} \partial_{\nu '}u) D^{\mu}F
      +(D^{\mu} \partial_{\nu'} u) D_{\mu} \partial_{\nu}F] 
      &  = 0 \, . \cr}
\eqno (2.19)
$$
This can be simplified using (2.16) and (2.17) to give 
$$
\eqalign{
    -(\partial_{\nu} \partial_{\nu'}u) [D^{\mu} \partial_{\mu}F
      + \xi (d+1)F + (\xi -1) (1+u)F'] \qquad& \cr
   {} - (\partial_{\nu} u \partial_{\nu'}u) [(\xi (d+2)-d)F'
       + (\xi -1) (1+u) F''] & 
       =0 \, .\cr}
\eqno (2.20)
$$ 
The coefficients of the two independent bitensors must each
vanish.  For the $\partial_{\nu} \partial_{\nu'}u$ term, we can
see that there is no choice of $\xi$ for which $F(u)$
obeys (2.13) as already established.  Thus there is no
value of the gauge parameter for which $S(u)$ vanishes. So we
proceed to the second stage in which we substitute the $ S(u)$
term of (2.8) in (2.5). This leads, after
similar algebra, to the superposition of bitensors
$$
\eqalign{ 
\xi (\partial_{\nu} \partial_{\nu'}u)[D^{\mu} \partial_{\mu} S'
     + 2 (1 + u) S'' +(d+1)S']  \qquad& \cr
  {}+\xi \partial_{\nu} u \partial_{\nu'}u [u(2+u) S'''' +
  (d+5) (1+u)S''' + 2 (d+2)S''] & \cr}
\eqno (2.21)
$$
which must be added to (2.20).

\medskip

We now choose Feynman gauge $\xi=1$ to simplify the equations.
Looking at the combined $\partial_{\nu} \partial_{\nu'}u$ term,
we impose an inhomogeneous condition on $S$ which will give the
desired equation (2.13) for $F$, namely
$$\eqalignno{
  \square S' + 2(1+u) S'' + (d+1) S' &= 2F & 
 (2.22)
\cr\noalign{or}
  u(2+u)S''' + (d+3)(1+u)S'' + (d+1)S' &= 2F
& (2.23)\cr}
$$
after use of (2.16).  The combined $\partial_{\nu} u
\partial_{\nu'}u$ bitensor term of (2.20) and (2.21) 
imposes another relation between $S$ and $F$ which must be
compatible with (2.23), and one can check by direct
differentiation of (2.23) that this is the case.

\medskip

Our program is now logically complete.  Eq. (2.22) is an 
inhomogeneous hypergeometric equation which is straightforward to 
solve.  We shall be content to check two details of the
solution.  First we note that as $u \to 0$, the inhomogeneous
solution of (2.22) behaves like $S'(u) \sim
1/u^{(d-3)/2}$ .  This is less singular than $F(u)$, so the gauge
term of (2.8) does not affect the short distance
behavior of the propagator.  As $u \to \infty$, one has,
analogously, $S'(u) \sim 1/u^{d-1}$.  We will need this information 
in the next section to check that possible surface terms from
$S(u)$ actually vanish.

\bigskip

\centerline{{\bf 3. Gauge Boson Exchange Integrals}}

\medskip
\noindent
We assume a schematic form of the AdS/CFT correspondence in which
a charged bulk scalar field $\varphi_\Delta (z)$ of mass $m$ is
the source for the scalar operator $\O_\Delta (x)$ in the
boundary theory, and $\Delta = \half \left( d + \sqrt{d^2 +
    4 m^2} \right)$.  Since we are interested only in the gauge boson
exchange process we can restrict our attention to the following terms of
the bulk action, namely
$$
  S _\varphi = \int d^{d+1}z \, \sqrt{g} 
      \left[ g^{\mu \nu} D_\mu \varphi^*_\Delta D_\nu
        \varphi_\Delta 
        + m^2 \varphi^*_\Delta \varphi_\Delta
      \right] + (\Delta \to \Delta ')
\eqno (3.1)
$$
with $D_\mu = \partial_\mu + i A_\mu (z)$, plus $S_A$ of (2.4).  The gauge group
of the bulk theory is $U (1)$, but results for non-abelian groups can easily
be attained from ours by supplying suitable group theory factors.

\medskip

\noindent
{\it a. Ingredients of the Witten diagram of Fig.\thinspace1 } 

\item{(1)}
  The bulk-to-boundary propagators [3]
 $$
   K_\Delta (z_0, \vec{z}, \vec{x}) 
       = C_\Delta \Bigl( 
                    {z_0 \over z^2_0 + \left(\vec{z} - \vec{x} \right)^2}
                  \Bigr)^\Delta
 \eqno (3.2)
$$
 with [15]
$$
C_\Delta =  {\Gamma (\Delta) \over \pi^{d/2} \Gamma (\Delta - d/2)}
\quad {\rm for} \quad \Delta > d/2, \qquad \qquad 
C_{d/2} =  { \Gamma (d/2) \over 2 \pi^{d/2}} 
\eqno (3.3)
$$

\item{(2)}
  The bulk-to-bulk photon propagator defined in (2.8)
  and (2.18). 

\item{(3)}
  An anti-symmetric derivative  at each $A_\mu \varphi^*
  \leftrightharpoon{\partial}_\mu \varphi$ vertex.

We factor out the normalization constants and define the
amplitude to be studied by
$$
  \bigl\langle \O_\Delta (x_1) \O_{\Delta'} (x_2)
     \O^*_\Delta (x_3) \O^*_{\Delta'} (x_4) \bigr\rangle _{\rm gauge}
   = - {\Gamma \left({d-1 \over 2} \right) \over 4 \pi^{(d+1) / 2} }
     C^2_\Delta C^2_{\Delta'} A (x_1, x_2, x_3, x_4)
\eqno (3.4)
$$
with
$$
\eqalign{
    A (x_i) &= 
      \int {d^{d+1}z \, d^{d+1}w \over z^{d-1}_0 \, w^{d-1}_0}
      \Bigl( {z_0 \over (z - x_1)^2} \Bigr)^\Delta
      \leftrightharpoon{\partial}_{\mu}
      \Bigl({z_0 \over (z - x_3)^2} \Bigr)^\Delta \cr
      &
 \quad {} \cdot    
      {\partial_\mu \partial_{\nu'} u \over (u (u+2))^{(d-1)/2}}
      \Bigl( {w_0 \over (w - x_2)^2} \Bigr)^{\Delta'}
      \leftrightharpoon{\partial}_{\nu'}
      \Bigl( {w_0 \over (w - x_4)^2} \Bigr)^{\Delta'}
\cr}
\eqno (3.5)
$$
where $\partial_\mu = {\partial \over \partial z_\mu}$ and
$\partial_{\nu'} = {\partial \over \partial w_{\nu'}}$.  To be
clear about the notation, we recall that $ (z - x_i)^2 = z^2_0  +
(\vec{z} - \vec{x}_i)^2$ and write the longer form on the right
hand side only when appropriate to emphasize the special role of
the $z_0$ coordinate.

In (3.5), we have dropped the $\partial_\mu
\partial_{\nu'} S(u)$ gauge term in the gauge boson
propagator (2.8) because the bulk contribution vanishes
upon partial integration due to current conservation.  However
there is also a surface term of the form
$$
  \lim_{z_0 \to 0} \int d^d z \, z^{-d+1}_0
  \Bigl({z_0 \over z^2_0 + (\vec{z} - \vec{x}_1)^2} \Bigr)^\Delta
  \leftrightharpoon{\partial}_{0}
  \Bigl({z_0 \over z^2_0 + (\vec{z} - \vec{x}_3)^2} \Bigr)^\Delta
  \partial_{\nu'} S(u)
\eqno (3.6)
$$
which must be shown to vanish.  To do this we use the long
distance behavior of $S' (u)$ established at the end of \S 2
and also (2.7).  One can then see that (3.6) is 
a difference of two integrals of the form
$$
  \lim_{z_0 \to 0} z^{2 \Delta}_0 \int d^d z \,
  {1 \over \bigl( z^2_0 + (\vec{z} - \vec{x}_1)^2
    \bigr)^{\Delta+1}}
  {1 \over \bigl( z^2_0 + (z - x_3)^2 \bigr)^\Delta}
  {N_{\nu'} \over \bigl( (z_0 - w_0)^2 + (\bar{z} - \bar{w})^2 \bigr)^{d-1}}
\eqno (3.7)
$$
where $N_{\nu'}$ has a non-vanishing smooth limit as $z_0 \to
0$.  By standard analysis of the possible singularities of the
integral in the limit, one can show that the limit vanishes if $d >
2$ and $\Delta \ge d/2$, as is our case.

\medskip

To prepare the way to calculate the integrals in $A (x_1, x_2,
x_3, x_4)$, we observe that the derivatives of the scalar
propagators produce terms of the form
$$
  (z - x_i)_\mu (\partial_\mu \partial_{\nu'} u) (w - x_j)_{\nu'} 
  = -{1 \over 2 z_0 w_0} \left [ (z - x_i)^2 + (w - x_j)^2 - (x_i - x_j)^2
                \right ]
\eqno (3.8)
$$
where we have used (2.6). 

\medskip

The first step is to simplify the integral by setting $x_1 = 0$,
and by changing integration variables using the inversion isometry of AdS,
namely $z_\mu = z'_\mu / (z')^2$ and $w_\mu = w'_\mu / (w')^2$,
with boundary points $x_2,x_3,x_4$ referred to their inverses by $x_i =
x'_i / (x_i ')^2$.  This method was clearly described
in [15].  The current at the $z$ vertex becomes
$$
\eqalign{
      J_\mu (z, x_1 = 0, x_3) &=
          \Bigl( {z_0 \over z^2} \Bigl)^\Delta
          \leftrightharpoon{\partial}_\mu
          \Bigl({z_0 \over (z - x_3)^2} \Bigr)^\Delta \cr
      &= - 2 \Delta (z')^2 J_{\mu \nu} (z') (x'_3)^{2 \Delta}
            (z_0')^{2 \Delta} 
          {(z' - x'_3)_\nu \over (z' - \vec{x}_{3}')^{2 (\Delta + 1)}}
\cr}
\eqno (3.9)
$$
where $z'^2 J_{\mu \nu} (z') = z'^2 \delta_{\mu \nu} - 2 z'_\mu
z'_\nu$ is the conformal jacobian [15].  The variable $u$
is inversion invariant, see (2.3), and the contraction $
J_\mu g^{\mu \nu} \partial_\nu \partial_{\nu'} u$ in (3.5)
is invariant, so the jacobian factor $J_{\mu \nu}$ cancels.  Similar remarks
apply to jacobian factors in $w'$.  The net result of the change of
variables is
$$
  A (x_1, x_2, x_3, x_4) = 2^d \Delta \Delta' 
        |x'_3|^{2 \Delta} |x'_2|^{2 \Delta'}
        |x'_4|^{2 \Delta'}
        B (x_1, x_2, x_3, x_4)
\eqno (3.10)
$$
with
$$
\eqalignno{
  B (x_i) 
  =& 
     \int {d^{d+1} w \ w_0^{2 \Delta' -1} \over 
         (w - x'_2)^{2 \Delta'} (w -x'_4)^{2 \Delta'}} 
     \cdot
      \int {d^{d+1} z \ z_0^{2 \Delta - 1} \over (z - x'_3)^{2
         \Delta + 1}} 
     \cdot
     {1 \over \left[ (z - w)^2 (z - w^*)^2 \right]^{(d-1) / 2}} 
     \cr
  & \quad {} \cdot
     \Bigl\{ {(z - x'_3)^2 - (x'_3 - x'_2)^2 \over (w - x'_2)^2}
        - {(z - x'_3)^2 - (x'_3 - x'_4)^2 \over (w - x'_4)^2}
      \Bigr\} &(3.11)
\cr}
$$
where we have dropped the primes on the integration variables $w, z$ and have
used $2 + u = (z -  w^*)^2 / 2 z_0 w_0$ with $w^*_\mu = (- w_0, \vec{w})$.

\medskip

\noindent
{\it b. Integrals over the interaction point $z$}

\medskip

We now study the $z$-integrals which take the form 
$$
  R_{\Delta, p} = \int {d^{d+1} z \, z^{2 \Delta -
      1}_0 \over \left[ (z - w)^2 (z - w^*)^2 \right]^{(d+1)/2}}
  {1 \over (z - x_3 ')^{2 p}}
\eqno (3.12)
$$
for the cases $p = \Delta, \Delta + 1$.  Our strategy is to use a
power series expansion in the product $z_0 w_0$ from the gauge boson
propagator denominators, to integrate, and then to resum the series. We
start with
$$
\eqalign{
    {1 \over \left[ (z - w)^2 (z - w^*)^2 \right ]^{(d-1)/2}}
    &= {1 \over \left[ z^2_0 + w^2_0 + (\vec{z} - \vec{w})^2 \right ]^{(d-1)}}
        {1 \over (1 - Y^2)^{(d-1)/2}}
        \cr
    &= {1 \over \left[ z^2_0 + w^2_0 + (\vec{z} - \vec{w})^2 \right]^{(d-1)}}
        \sum^\infty_{k = 0}
        {\Gamma \left( k + {d-1 \over 2} \right) \over \Gamma
          \left( {d-1 \over 2} \right) k!}
        Y^{2 k} 
\cr}
\eqno (3.13)
$$
with $Y \equiv 2 z_0 w_0 / \left( z^2_0 + w^2_0 + (\vec{z} -
  \vec{w})^2 \right)$.  Note that $Y \leq 1$. 
We then have the convergent series expansion 
$$
  R_{\Delta, p} = \sum^\infty_{k = 0}
     {\Gamma \left( k + {d-1 \over 2} \right) \over \Gamma \left(
        {d-1 \over 2} \right) k!}
     (4 w^2_0)^k R_k
\eqno (3.14)
$$
with the integrals
$$
  R_k = \int^\infty_0 d z_0 ~ z_0^{2 (\Delta + k) -1}
        \int d^d z 
        {1 \over \left[ w^2_0 + z^2_0 + (\vec{z} - \vec{w})^2
          \right]^{d - 1 + 2k}
          \left( z^2_0 + (\vec{z} - \vec{x}'_3)^2 \right)^p}
\eqno (3.15)
$$
The spatial integral in (3.15) can be done by introducing one Feynman parameter
$\alpha$, with the result
$$
  {\pi^{d/2} \Gamma (\sigma) \over \Gamma (d - 1 + 2k)
    \Gamma (p)}
  \int^1_0
  {d \alpha \, \alpha^{d - 2 + 2k} (1 -
    \alpha)^{p-1} \over \left[ z^2_0 + \alpha (1 - \alpha) 
      (\vec{w} - \vec{x}'_3)^2 + \alpha w^2_0 \right]^\sigma}
\eqno (3.16)
$$
where we use the abbreviation $\sigma = 2 k + p - 1 + d/2$.  The $z_0$ integral
is then straightforward,
$$
  \int^\infty_0 {d z_0 ~ z_0^{2 (\Delta + k) - 1} \over \left[
      z^2_0 + \alpha (1 - \alpha) (\vec{w} - \vec{x}_3 ')^2 +
      \alpha w^2_0 \right]^\sigma}
  = {\Gamma (\Delta + k) \Gamma (\sigma - \Delta - k)
 \alpha^{ k + \Delta -\sigma}\over 2
    \Gamma (\sigma) 
    \left[ w^2_0 + (1 - \alpha) (\vec{w} - x'_3) \right]^{\sigma - \Delta -k}}
\eqno (3.17)
$$
Putting things together and applying the doubling formula to
$\Gamma (d - 1 + 2k)$, we finally obtain
$$
\eqalign{
    R_{\Delta, p} &=
       {\pi^{(d + 1)/2} \over 2^{d - 1}}
       {1 \over \Gamma (p) \Gamma \left( {d - 1 \over 2} \right)}
       \sum^\infty_ {k=0}
       {\Gamma (\Delta + k) \Gamma \left({d \over 2} + k +
           p - \Delta - 1 \right) w^{2k}_0 \over \Gamma \left( k +
           {d \over 2} \right) k!}
           \cr
   & \quad {} \cdot
       \int^1_0 d \alpha
       {\alpha^{{d \over 2} + k + \Delta - p - 1} (1 -
         \alpha)^{p - 1} \over \left [ w^2_0 + (1 - \alpha) (\vec{w} -
           \vec{x}'_3)^2\right]^{{d \over 2} + k - 1 + p - \Delta}}
\cr}
\eqno (3.18)
$$
The resummation of the series is simplest in the case $p = \Delta 
+ 1$.  There is a cancellation of $\Gamma$ functions, and we can
recognize the binomial series
$$
  \sum^\infty_{k - 0} 
  {\Gamma (\Delta + k) \over \Gamma (\Delta) k!}
  x^k = {1 \over (1 - x)^\Delta}
\eqno (3.19)
$$
in the variable $x = \alpha w^2_0 / [w^2_0 + (1 - \alpha)
(\vec{w} - \vec{x}_3 ')^2]$.  We thus obtain the exact formula
$$
\eqalign{
    R_{\Delta, \Delta + 1} &=
        {\pi^{(d + 1) / 2} \over 2^{d - 1} \Delta \Gamma 
            \left( {d - 1 \over 2} \right)}
        {1 \over \left[ w^2_0 + (\vec{w} - \vec{x}'_3)^2 \right]^\Delta}
        \cr
    & \quad {} \cdot
        \int^1_0 d \alpha \, \alpha^{{d \over 2} - 2}
          \left[ w^2_0 + (1 - \alpha) (\vec{w} - \vec{x}'_3)^2
          \right]^{\Delta - d/2}
\cr}
\eqno (3.20)
$$
Direct resummation of the series for
$R_{\Delta, \Delta}$ would be more difficult, but it can be
avoided if we first integrate by parts term-by-term using $(1 -
\alpha)^{\Delta - 1} = - (1 / \Delta) {d \over d \alpha} (1 -
\alpha)^\Delta$.  We then quickly derive the general relation
$$
  R_{\Delta, \Delta} = 
    \left[ w^2_0 + (\vec{w} - \vec{x}'_3)^2 \right]
    R_{\Delta, \Delta + 1} \, .
\eqno (3.21)
$$
Note that when $d$ is even and $\Delta$ an integer satisfying the 
unitarity bound $\Delta \geq d/2$, the integrand of (3.20) is just a simple
polynomial in~$\alpha$ and the integral can be evaluated as a finite sum of 
elementary terms.  
$$
\eqalign{
  R_{\Delta, \Delta + 1} 
= {\pi^{{d+1 \over 2}} \over 2^{d-1} \Gamma ({d-1 \over 2}) } &
     {\Gamma (\Delta - {d \over 2} +1) \over \Gamma (\Delta +1)}
     {1 \over \left( w^2_0 + (\vec{w} - \vec{x}'_3)^2 \right)^{d/2}} \cr
    & \cdot \sum^{\Delta - {d\over 2}}_{\ell = 0}
     {\Gamma (\ell + {d\over 2} -1) \over \ell !}
     \Bigl({w^2_0 \over w^2_0 + (\vec{w} - \vec{x}'_3)^2} \Bigr)^\ell
\cr}
\eqno (3.22)
$$
The restriction to even $d$ and integer $\Delta$ includes the $d=4$, $\N=4$ 
super--Yang-Mills theory which has chiral primary operators of integer dimension.
Our current progress may be summarized by inserting
(3.21) and (3.22) into (3.11) which now reads
$$
      B (x_i)  
      = 
      \int {d^{d + 1} w \ w _0 ^{2 \Delta' - 1} R_{\Delta, \Delta
         + 1} \over (w - x'_2)^{2 \Delta'} (w - x'_4)^{2 \Delta'}}
      \Bigl[ {(w - x'_3)^2 - (x_3 ' - x'_2)^2 \over (w - x'_2)^2}
        - {(w - x'_3)^2 - (x'_3 - x'_4)^2 \over (w - x'_4)^2}
      \Bigr]
\eqno (3.23)
$$
It remains to carry out the integrals over the interaction point $w$.

\medskip

\noindent
{\it c. Integrals over the interaction point $w$}

\medskip

We see that the remaining integrals over $w$ are of the form
$$
  S_k ^{(\ell)} 
= \int {d^{d + 1} w \ w_0 ^{2 \Delta' - 1} \over (w - \vec{x})^{2
      \Delta'} (w - \vec{y})^{2 (\Delta' + 1) }} ~
     {1 \over (w^2)^{d/2-2+k}} ~
     \Bigl( {w^2_0 \over w^2}\Bigr)^\ell
\eqno (3.24)
$$
for the two cases $k = 1$ and $k=2$, and with $x \equiv x'_4 - x'_3$, $y
\equiv x'_2 - x'_3$ minus the reverse assignment of $x, y$ in (3.23). 
Specifically, we have
$$
\eqalign{
B(x_i) 
  & = {\pi^{{d+1 \over 2}} \over 2^{d-1} \Gamma ({d-1 \over 2}) } 
     {\Gamma (\Delta - {d \over 2} +1) \over \Gamma (\Delta +1)}
     \sum^{\Delta - {d\over 2}}_{\ell = 0}
     {\Gamma (\ell + {d\over 2} -1) \over \ell !} S ^{(\ell)} \cr
S^{(\ell)} 
  &=    S_1 ^{(\ell)} - y^2 S_2 ^{(\ell)} 
       -(x \leftrightarrow y)  
\cr}
\eqno (3.25)
$$
The integral can now be done by the following standard steps:
  (a) combine $w^2$ and $(w-x)^2$ denominators with Feynman parameter
      $\alpha$,
  (b) combine the composite denominator from the previous step with the
      $(w-y)^2$
      denominator using Feynman parameter $\beta$,
  (c) carry out the $d^dw$ integral,
  (d) do $dw_0$ integral as in (3.16). 
We suppress details and directly give the result
$$
\eqalign{
    S_k ^{(\ell)} &= {\pi^{d/2} \Gamma (\Delta' + \ell) \Gamma 
                  (\Delta' + k -1) \over 
                2 \Gamma (\ell + k +d/2 -2 ) \Gamma (\Delta') \Gamma
                  (\Delta' + 1)}
              \cr
        & \quad {} \cdot
           \int^1_0 d \alpha
           \int^1_0 d \beta \
           {\alpha^{\Delta' - 1} (1 - \alpha)^{\ell + {d\over 2} + k - 3 }
                \beta^{\Delta'} (1-\beta)^{\ell + {d \over 2} - 2} 
             \over \left[ \beta (y - \alpha x)^2 + \alpha (1 -
                 \alpha) x^2 \right]^
               {\Delta' + k - 1 }}.
\cr}
\eqno (3.26)
$$
The above representation is not well-defined when $\ell + d/2-1=0$. However, we
shall be most interested in the cases with $d \geq 3$, where this special
configuration does not occur, and we shall henceforth assume that $d \geq 3$,
so that $\ell + d/2 -1 >0$. 

\medskip

\noindent
{\it d. Combining all integrals}

\medskip

An improved integral representation may be obtained in which the four
$S_k ^{(\ell)}$ contributions to $S^{(\ell)}$ are combined in a single
expression which directly exhibits the natural anti-symmetry of the amplitude
under the interchange of $x$ and $y$. This new form will prove very useful in
obtaining logarithmic and power-singular contributions to the amplitude in 
various
channels, as well as to obtaining a complete OPE expansion of the amplitude. 
First, we carry out a change of variables :
$\beta = 1/(1+\xi)$, so that
$$
\int _0 ^1  \!\!\! 
  { d \beta \ \beta ^{\Delta '} (1 - \beta )^{\ell +\d -2}
  \over
  [\beta (y-\alpha x)^2 + \alpha (1 - \alpha ) x^2 ] ^{\Delta ' +k -1} }
=
\int _0 ^\infty \!\!\! \!
  { d \xi \ \xi ^{\ell +\d -2} (1 + \xi )^{-\ell -\d +k -1} 
  \over
  [y^2 -2 \alpha x\cdot y + \alpha x^2 + \xi \alpha (1-\alpha) x^2 ]^{\Delta '
   +k -1} } \, .
\eqno (3.27)
$$
The exponents in the denominators for $k=1,2$ differ by one, but we may render
these identical by integrating by parts in the variable $\xi$ for the $k=1$
integral and using 
$$
{d \over d\xi} 
{\xi ^{\ell +\d -1} \over (1 +\xi) ^{\ell +\d -1} }
= 
(\ell +\d -1) {\xi ^{\ell +\d -2} \over (1 +\xi) ^{\ell +\d } } \, .
$$
It is now easy to combine the $k=1,2$ contributions, and we obtain
$$
\eqalignno{
S_1 ^{(\ell)}  - y^2 S_2 ^{(\ell)}
  &= {\pi^{d/2} \Gamma (\Delta' + \ell)  \over 
                2 \Gamma (\ell +\d ) \Gamma (\Delta') }
      & (3.28)       \cr
        & \quad {} \cdot
           \int^1_0 d \alpha
           \int^\infty _0 d \xi \
           {\alpha^{\Delta' - 1} (1 - \alpha)^{\ell + \d - 1 }
                \xi^{\ell +\d -2} (1+\xi)^{-\ell - \d +1} (\alpha \xi x^2 -y^2)
             \over \left[ y^2 - 2 \alpha x\cdot y + \alpha x^2 + \xi \alpha (1 -
                 \alpha) x^2 \right]^
               {\Delta' + 1 }}
\cr}
$$
We now perform three consecutive changes of variables : 
\smallskip
\item{(a)} let $\eta \equiv \xi \alpha (1-\alpha)$;
\smallskip
\item{(b)} let $\alpha \equiv 1/(1+u)$,
\smallskip
\item{(c)} let $v \equiv \eta (1+u)$.
\smallskip
\noindent
We find that the combination $S_1 ^{(\ell)}  - y^2 S_2 ^{(\ell)}$ 
now automatically changes sign under the interchange of $x$ and $y$, so that
we have $S^{(\ell)} = 2(S_1 ^{(\ell)}  - y^2 S_2 ^{(\ell)})$. The final
result is the following simple integral representation 
$$
S ^{(\ell)} 
  = {\pi^{d/2} \Gamma (\Delta' + \ell)  \over 
                 \Gamma (\ell +\d ) \Gamma (\Delta') }
          \int^\infty _0 \! \! \! du
           \int^\infty _0 \! \! \!  dv \
           {(uv)^{\ell + \d - 2 } \over (u + v + uv) ^{\ell +\d -1} } \cdot
           {vx^2 - u y^2  \over \left[u y^2 +(x-y)^2 + v x ^2 \right]^
               {\Delta' + 1 }}
\eqno (3.29)
$$
This double integral representation is a convenient starting point for the
study of the amplitude in various limits, as will be carried out in the next
section.

\bigskip

\centerline{\bf 4. Explicit formulas and Singularity Structure}

\medskip

Combining all results of the preceding section for the amplitude $A(x_i)$, we
may express the reduced form $B(x_i)$, as defined in (3.10), in terms of the
following conformal invariants
$$
s \equiv \half {(x-y)^2 \over x^2 + y^2}
\qquad {\rm and } \qquad
t \equiv {x^2 - y^2 \over x^2 + y^2} \, .
\eqno (4.1\rm a)
$$
In terms of the original position coordinates $x_i$ of the amplitude, these
variables take the form
$$
s= \half { x_{13} ^2 x_{24}^2 \over x_{12}^2 x_{34}^2 + x_{14}^2 x_{23}^2 }
\qquad \qquad
t= {x_{12}^2 x_{34}^2 - x_{14}^2 x_{23}^2 \over x_{12}^2 x_{34}^2 + x_{14}^2
x_{23}^2 }.
\eqno (4.1\rm b)
$$ 
As $x$ and $y$ vary, we have $0\leq s\leq 1$ and $-1 \leq t \leq 1$. It is
convenient to express $B$ as follows
$$
B(x_i) 
  = {\pi ^{d+\half} \Gamma (\Delta - \d +1) \over 2^{d-1} \Gamma (\Delta ')
    \Gamma (\d - \half) \Gamma (\Delta +1) (x^2 + y^2 )^{\Delta '} }
    \sum _{\ell =0} ^{\Delta -\d} 
    {\Gamma (\Delta ' + \ell) \Gamma (\ell + \d -1) \over
     \Gamma (\ell + \d) \ell !} I^{(\ell)}
\eqno (4.2)
$$
where $I^{(\ell)}$ is given in terms of $s$ and $t$ only :
$$
I^{(\ell)} = \int^\infty _0 \! \! \! du
           \int^\infty _0 \! \! \!  dv \
           {(uv)^{\ell + \d - 2 } \over (u + v + uv) ^{\ell +\d -1} } \cdot
           {\half (u+v)t  +\half (v-u)  \over \left[\half (u+v) +\half (v-u)t + 
2
s
\right]^
               {\Delta' + 1 }}.
\eqno (4.3)
$$
We notice right away that the integrals for different values of $\Delta'$ are
related to one another by differentiation with respect to the parameter
$s$. Actually, the integrals for different values of $\ell$ (and in fact also
different $d$) are also related to one another, but to see this easily, it is
convenient to perform one penultimate change of variables : $u= 2\rho
(1-\lambda)$ and $v=2 \rho (1+\lambda)$, so that
$$
I^{(\ell)} = 2^{1-\Delta '} \int^\infty _0 \! \! \! d\rho
           \int^{+1} _{-1} \! \! \!  d\lambda \
           {\rho ^{\ell +\d -1} (1-\lambda ^2)^{\ell + \d - 2 }  
   \over [1 + \rho (1 - \lambda ^2)] ^{\ell +\d -1} }
\cdot
           {\lambda + t  \over \left[ \rho + \rho \lambda t +s \right]^
               {\Delta' + 1 }} \, .
\eqno (4.4)
$$
We now make use of the elementary formula
$$
{1 \over (s + a) ^k} 
  = { (-)^k \over \Gamma (k) }
   \Bigl ( {\partial \over \partial s} \Bigr ) ^{k-2} 
   {1 \over (s +a) ^2}
= { (-)^{k+1} \over \Gamma (k) }
   \Bigl ( {\partial \over \partial s} \Bigr ) ^{k-1} 
   {1 \over s +a}
\eqno (4.5)
$$
and obtain for $k=\Delta '+1$
$$
I^{(\ell)} = {(-2)^{1-\Delta '} \over \Gamma (\Delta '+1)}
             \Bigl ( {\partial \over \partial s} \Bigr ) ^{\Delta '-1}
              \int^\infty _0 \! \! \! d\rho
           \int^{+1} _{-1} \! \! \!  d\lambda \
           {\rho ^{\ell +\d -1} (1-\lambda ^2)^{\ell + \d - 2 }  
   \over [1 + \rho (1 - \lambda ^2)] ^{\ell +\d -1} }
\cdot
           {\lambda + t  \over \left[ \rho + \rho \lambda t +s \right]^ 2} \, .
\eqno (4.6)
$$
Letting now $\mu \equiv s/\rho$, and noticing that by using (4.5) again, we
have
$$
{ (1-\lambda ^2) ^{\ell +\d -2}
  \over 
  [\mu + s (1-\lambda ^2)] ^{\ell +\d -1}}
= { (-)^{\ell +\d} \over \Gamma (\ell + \d -1) }
 \Bigl ( {\partial \over \partial s} \Bigr ) ^{\ell +\d -2}
{1 \over \mu + s (1-\lambda ^2)}
\eqno (4.7)
$$
we obtain our final formula for the amplitudes, in which all $I^{(\ell)}$ are
expressed as derivatives of a single universal function $I$, which is
independent
of $d$, $\Delta '$ and $\ell$.
$$
\eqalign{
I^{(\ell)} & = {2^{1-\Delta '} (-) ^{\ell +\d + \Delta ' -1}
             \over \Gamma (\Delta'+1) \Gamma (\ell +\d -1)}
             \Bigl ( {\partial \over \partial s} \Bigr ) ^{\Delta '-1}
              \Bigl \{ s^{\ell +\d -2} 
             \Bigl ( {\partial \over \partial s} \Bigr ) ^{\ell +\d -2}
             I (s,t) \Bigr \} 
\cr
I (s,t)    &=  \int^\infty _0 \! \! \! d\mu
           \int^{+1} _{-1} \! \! \!  d\lambda \
           {1 \over \mu + s (1 - \lambda ^2)  }
\cdot
           {\lambda + t  \over ( \mu +  \lambda t +1)^ 2} \, .
\cr}
\eqno (4.8)
$$
The $\mu$-integral is elementary and may be carried out right away. One obtains
$$
\eqalign{
I(s,t) 
  = \int _{-1} ^{+1} \! \! \! d\lambda &
     {\lambda + t \over [1+\lambda t - s (1-\lambda ^2)]^2}
     \ln \Bigl ({1+\lambda t \over s (1-\lambda ^2)} \Bigr )
     \cr
  &- \int _{-1} ^{+1} \! \! \! d\lambda 
     {\lambda + t \over 1+\lambda t - s (1-\lambda ^2)}
      \cdot {1 \over 1+ \lambda t} \, .
     \cr}
\eqno (4.9)
$$
The $\lambda$ integration of the second term and the term proportional to $\ln
s$ in the first term are still elementary, but the other contributions in the
first term involve dilog or Spence functions and are not elementary.

\medbreak

\noindent
{\it a. The expansion in the direct channel : $x_2 - x_4 \to 0$}

\medskip

The singularities and OPE expansion in the gauge boson channel are obtained
from an expansion for small $x_2' - x_4'$, which corresponds to $x-y$ small,
and thus to $s $ and $t$, defined in (4.1), tending to 0. This systematic
expansion may be directly read off from the integral representation in (4.9),
upon series expansion in powers of $s$ and $t$. Now, for $d\geq
4$ -- which we shall henceforth restrict to -- the power factor of $s$
between the two derivative factors in (4.8) cannot cause a singularity. Thus,
all non-analyticity of $B$ must arise from the non-analytic part of $I(s,t)$
as $s\to 0$, which we denote here by $I_{{\rm sing}} (s,t)$. It is instructive
to set
$$
I(s,t) = I_{{\rm sing}} (s,t) + I_{{\rm reg}} (s,t)
\eqno (4.10)
$$
where $I_{{\rm reg}} (s,t)$ now admits a Taylor series expansion in powers of
$s$.

We first examine the non-analytic part $I_{{\rm sing}} (s,t)$ which is
proportional to the explicit $\ln s$ dependence in (4.9). We give this part
below as well as the full answer of the integral involved 
$$
\eqalignno{
I_{{\rm sing}} (s,t) 
  & = - \ln s \cdot \int _{-1} ^{+1} \! \! \! d\lambda
   { \lambda + t \over [1+ \lambda t - s (1-\lambda ^2)]^2} &(4.11)\cr
  & = {- \ \ln s \over s^2  (\lambda _+ - \lambda _-)^2}
     \Bigl \{ 
       {\lambda _+ + \lambda _- + 2t \over \lambda _+ - \lambda _-}
       \ln {(1-\lambda _-) (1+ \lambda _+) \over (1-\lambda _+) (1+ \lambda _-)}
            -2 {\lambda _+ + t \over 1 - \lambda _+ ^2}
            -2 {\lambda _- + t \over 1 - \lambda _- ^2}
      \Bigr \}
\cr}
$$
where $\lambda _\pm$ are defined by
$$
\lambda _\pm = - {t \over 2 s} \pm
\sqrt{ {t^2 \over 4 s^2} +1 - {1 \over s}} \, .
\eqno (4.12)
$$
A systematic expansion in powers of $s$ is given by the following series
$$
\eqalign{
I_{{\rm sing}} (s,t) 
   & = \ln s \sum _{k=0} ^ \infty (k+1)\, s^{k}\, a_k(t) \cr
a_k(t) 
   & = \int _{-1} ^{+1} \! \! \! d\lambda \
   { (\lambda + t) (1-\lambda ^2) ^k  \over (1+ \lambda t )^{k+2}} \, .
\cr}
\eqno (4.13)
$$
The function $a_k(t)$ is hypergeometric and admits the following series 
expansion in powers of $t$
$$
\eqalign{
a_k (t) & =
  2^k {\Gamma (\k +1) \Gamma (\k +{3 \over 2}) \over \Gamma (k +{5 \over 2})}
  \, t \, F(\k +1, \k +{3\over 2}; k +{5 \over 2}; t^2 )
\cr
        & =
  \sqrt \pi  \sum _{m=0} ^\infty 
    {\Gamma (2m+k+2) \over \Gamma (m+k+{5 \over 2})\,  m!} 
    \bigl ( {t\over 2} \bigr ) ^{2m+1} \, .
\cr}
\eqno (4.14)
$$
This result may be used to derive the singular and logarithmic parts of
$I^{(\ell)}$ as well as the associated analytic part. It is helpful to
make use of the following formulas
$$
\eqalign{
\Bigl ({\partial  \over \partial s } \Bigr) ^\ell s^k
 & = {\Gamma (k+1) \over \Gamma (k-\ell +1) } s^{k-\ell} 
\cr
\Bigl ( {\partial  \over \partial s } \Bigr ) ^\ell \{ s^k \ln s \} 
& = {\Gamma (k+1) \over
\Gamma (k-\ell +1) } s^{k-\ell} \{ \ln s + \psi (k+1) - \psi (k-\ell +1) \}
\cr}
\eqno (4.15)
$$
where $\psi (x) = \Gamma '(x) /\Gamma (x)$. We find
$$
\eqalignno{
I_{{\rm sing}} ^{(\ell)}
= & 
 {2^{1-\Delta '} (-) ^{\ell +\d + \Delta ' -1}
             \over \Gamma (\Delta'+1) \Gamma (\ell +\d -1)}
             \sum _{k=0} ^\infty {\Gamma (k+1) \Gamma (k+2) \over 
             \Gamma (k-\ell -\d +3) \Gamma (k-\Delta '+2)}
& (4.16)\cr
& {} \qquad \qquad 
             s^{k-\Delta ' +1} a_k(t) \Bigl \{
   \ln s + 2 \psi (k+1) - \psi (k-\ell-\d +3) - \psi (k -\Delta ' +2) \Bigr \}.
\cr}
$$
This expression is still somewhat formal, since the $\Gamma$-functions in the
denominator inside the sum may be infinite for sufficiently small $k$, while
the $\psi$-functions may diverge there. To properly separate out this behavior,
we introduce
$$
\eqalign{
k_+ & = {\rm max} \{ \Delta ' -1, \ \ell +\d -2\} \cr
k_- & = {\rm min} \{ \Delta ' -1, \ \ell +\d -2\}\, . \cr}
\eqno (4.17)
$$
When $k\geq k_+$, no divergences occur and the series has well-defined terms 
as it stands. For $k<k_-$, the divergences of both $\Gamma$ functions
produces a double zero, while $\psi$-functions only produce a single pole, so
that these contributions cancel. Thus, we have
$$
\eqalignno{
I_{{\rm sing}} ^{(\ell)}
= & - 
 {2^{1-\Delta '} (-) ^{\ell +\d + \Delta ' -1}
             \over \Gamma (\Delta'+1) \Gamma (\ell +\d -1)}
    \sum _{k= k_-} ^{k_+ -1}{\Gamma (k+1) \Gamma (k+2)\psi (k+1-k_+)  \over 
              \Gamma (k-k_- +1) \Gamma (k+1-k_+)}  
 s^{k-\Delta '+1} a_k(t)
\cr
+ & 
 {2^{1-\Delta '} (-) ^{\ell +\d + \Delta ' -1}
             \over \Gamma (\Delta'+1) \Gamma (\ell +\d -1)}
             \sum _{k=k_+} ^\infty {\Gamma (k+1) \Gamma (k+2) \over 
             \Gamma (k-\ell -\d +3) \Gamma (k-\Delta '+2)}
& (4.18)\cr
& {} \qquad \qquad 
           \cdot  s^{k-\Delta ' +1}a_k(t) \Bigl \{
   \ln s + 2 \psi (k+1) - \psi (k-\ell-\d +3) - \psi (k -\Delta ' +2) \Bigr \}
\cr}
$$
When $k_- \leq k \leq k_+-1$, the ratio $\psi (k+1-k_+) /\Gamma (k+1-k_+)$ is a
finite number, and thus the first line of (4.18) is the most singular
contribution to $I_{{\rm sing}} ^{(\ell)}$. 
The leading logarithmic contribution enters when $k=k_+$, and takes the form
$$
I^{(\ell)} _{{\rm leading \ log}}
 = {2^{1-\Delta '} (-) ^{\ell +\d + \Delta ' -1}
             \over \Gamma (\Delta'+1) \Gamma (\ell +\d -1)}
{\Gamma (k_+ +1) \Gamma (k_+ +2) \over 
              \Gamma (k_+ -k_- +1)}   \ s^{k_+ -\Delta '+1} \ln s \ a_{k_+} (t)
\eqno (4.19)
$$

The regular part $I_{{\rm reg}} (s,t)$ admits a similar expansion in powers of
$s$, given as follows:
$$
\eqalign{
I_{{\rm reg}} (s,t) 
  &=
     \sum _{k=0} ^\infty s^k b_k (t) \cr
b_k (t) 
  &=
     \int _{-1} ^{+1} d \lambda 
      {(\lambda + t) (1 - \lambda ^2)^k \over (1+\lambda t)^{k+2}}
      \Bigl \{ -1 - \lambda t +(k+1)\ln {1+\lambda t \over 1-\lambda ^2} \Bigr 
\}.
\cr}
\eqno (4.20)
$$
The contribution of the analytic part may be inserted to obtain the regular
contribution to $I^{(\ell)}$, just as we did for $I_{{\rm sing}}$ above.

\medskip

It is now straightforward, but somewhat tedious, to assemble all results and to
extract the physical $t$-channel limit of the gauge boson exchange amplitude,
namely the limit in which $x_{13}$ and $x_{24}$ are small compared to
other coordinate differences such as $x_{12}$. In this limit the variable $t$
of (4.1b) can be rewritten as $ t \sim - x_{13} \cdot J(x_{12})\cdot x_{24}/x_{12}^2$,
where $J_{\mu\nu}(y) = \delta_{\mu\nu} - 2 y_\mu y_\nu/y^2$ is the well-known
Jacobian of the conformal inversion.

\medskip

To incorporate the $1/{(x^2+y^2)}^{\Delta'}$ factor in (4.2) we also need 
$$
\eqalign{
x^2 & = x_{34}^2/x_{14}^2 x_{13}^2 \sim 1/x_{13}^2   
\cr\noalign{\smallskip}
y^2 & = x_{23}^2/ x_{12}^2 x_{13}^2 \sim 1/x_{13}^2.
\cr}
\eqno (4.21)
$$
Finally we need the conformal inversion prefactor in (3.10), which is
$$
|x_3'|^{2\Delta} |x_2'|^{2\Delta'} |x_4'|^{2\Delta'} 
=
1/|x_{13}|^{2\Delta} |x_{12}|^{2\Delta'} |x_{14}|^{2\Delta'} 
\sim
1/|x_{13}|^{2\Delta} |x_{12}|^{4\Delta'}.
\eqno (4.22)
$$
The procedure is then to substitute $I^{(\ell)}_{{\rm sing}}$ from (4.18) into 
(4.2)
and then into (3.10) and finally (3.4). The extraction of a complete
asymptotic series in this limit requires non-leading corrections to the
kinematic equations above, so we will restrict our discussion to the
power dependence of leading terms. The procedure above gives the leading
power term
$$\eqalign{
 \bigl\langle \O_\Delta (x_1) \O_{\Delta'} (x_2)& 
     \O^*_\Delta (x_3) \O^*_{\Delta'} (x_4) \bigr\rangle _{\rm gauge}\cr
 &\sim x_{13} \cdot J(x_{12})\cdot x_{24}/ |x_{13}|^{2\Delta -d +2}
 |x_{12}|^{2(d-1)} |x_{24}|^{2\Delta' -d+2}.}
\eqno (4.23)
$$
The angular dependent numerator and the various exponents are exactly as
 expected for the contribution of a conserved current of dimension $d-1$
in the double OPE of the scalar operators in the correlator. Similarly, the
leading logarithmic contribution is given by the product of (last eq) with
the factor $s^{(\Delta'-d/2 +1)} \ln s$. Note that this discussion has assumed
that $\Delta' -1 > d/2+1$ in (4.17) and that the leading terms come from
$\ell =0$.

\medbreak

\noindent
{\it b. Expansion in the crossed channel : $x_3-x_4 \to 0$}

\medskip

A systematic expansion for $x_3 - x_4 \to 0$ may be derived in a way analogous 
to the expansion for $x_2 - x_4 \to 0$, carried out in \S a. above. We begin  
by obtaining the leading term as $x_3 - x_4 \to 0$, and thus as $x\to 0$. 
Remarkably, the leading term is most conveniently obtained directly 
from the original Feynman parameter integrals in (3.25) and (3.26). As it was 
shown in (3.28) and (3.29) that 
$S^{(\ell)} = 2 (S_1 ^{(\ell)} - y^2 S_2 ^{(\ell)}$, 
it suffices to analyze the asymptotics of $S_1 ^{(\ell)}$ and $S_2  ^{(\ell)}$ 
directly from (3.26). Concentrating on the $\beta$-integral as $x\to 
0$ first, we see that $S_2 ^{(\ell)}$ is dominated by a $\ln x^2$ term, while
$S_1 ^{(\ell)}$ is dominated by a term of order 1. Thus, the contribution of 
$S_1 ^{(\ell)}$ is negligible. Furthermore, using the asymptotics of the 
following integral
$$
\int _0 ^1 d \beta { \beta ^{\Delta '} \over (\beta +b)^{\Delta '+1}}
= - \ln b + \O (1)
\eqno (4.24)
$$
as $b \to 0$, the $\beta $-integral in (3.26) for $k=2$ becomes trivial, and the 
$\alpha$-integral reduces to a ratio of $\Gamma$-functions. One finds
$$
S^{(\ell)} 
 = { \pi ^\d \Gamma (\Delta '+\ell) \over \Gamma (\ell + \d + \Delta ') }
{1 \over y ^{2 \Delta '} }\ln { x^2 \over y^2}.
\eqno (4.25)
$$
This result means that for any $\Delta$, $\Delta '$ and $d$, the leading 
behavior of the amplitude in the crossed channel is given by a pure logarithm of 
$x^2$, with no power singularities.

\medskip

A systematic expansion for $x_{34} \to 0$ to all orders may be obtained from the 
universal function $I(s,t)$ in (4.8), just as for the case $x_{24} \to 0$. 
Clearly, from
(4.1a) the limit $x_{34} \to 0$ corresponds to $x\to 0$ and thus implies that 
$s \to \half$ and $t \to -1$. 
To carry out the expansion, we consider the double integral representation of
$I(s,t)$ in (4.8) and perform two consecutive changes of variables
\smallskip
\item{(a)} $\mu = \half (1-\lambda ^2) (\eta -1)$ with $1 \leq \eta$,
\smallskip
\item{(b)} $\lambda = (\xi -1)/(\xi +1)$ with $0\leq \xi \leq \infty$.
\smallskip

It is convenient to set  $t=-1 +2\tau$ at intermediate stages of the
calculation. The result is the following double integral representation
$$
I(s,t) 
= \int _1 ^\infty {d \eta \over \eta - (1-2s)}  \int _0 ^\infty d \xi 
 { \tau (\xi +1)^2 -\xi - 1 \over (\tau \xi ^2 + \eta \xi +1 -\tau)^2}.
\eqno (4.27)
$$
The $\xi$-integral is easily carried out, and equals
$$
{2-4\tau \over \eta ^2 -4\tau (1-\tau)}
+ {\eta (1-2\tau) 
  \over 
  \big [\eta ^2 - 4 \tau (1-\tau) \big ]^{3/2}}
  \Bigl \{ \ln \tau (1-\tau) - 2\ln  \big [{ \eta \over 2}  
  + \half \sqrt{\eta ^2 - 4 \tau (1-\tau)} \big ] \Bigr \}.
\eqno (4.28)
$$
The expression for $I(s,t)$ in (4.27) manifestly admits a convergent Taylor
series expansion in powers of $(2s-1)$,
$$
I(s,t) = \sum _{n =0} ^\infty  (1-2s)^n 
             \{ a^{(n)} _{{\rm sing}} (t) + a^{(n)} _{{\rm reg}} (t) \}.
\eqno (4.29)
$$
In view of (4.28), the coefficients are given by
$$
\eqalign{
a^{(n)} _{{\rm sing}} (t) 
& = -t \, \ln {1 -t^2 \over 4}   \int _1 ^\infty d\eta \,
    { \eta ^{-n} \over \big [ \eta ^2 - (1-t^2) \big]^{3/2} }
\cr
&\cr
a^{(n)} _{{\rm reg}} (t)
& = 2t \int _1 ^\infty d\eta \,
   {\eta ^{-n} \ \ln \big [ {\eta \over 2} + \half \sqrt {\eta ^2 - (1-t^2)} 
\big
] 
   \over 
   \big [ \eta ^2 - (1-t^2) \big ]^{3/2} }
-  
   2t \, \int _1 ^\infty d\eta \,
   {\eta ^{-n-1} 
   \over 
   \eta ^2 - (1-t^2)  }.
\cr}
\eqno (4.30)
$$
The functions $a^{(n)} _{{\rm sing}} (t)$ are proportional to a
hypergeometric function, as well as to a logarithm of $1-t^2$, 
$$
\eqalign{
a^{(n)} _{{\rm sing}} (t) 
&=
-t \, \ln {1 - t^2 \over 4}  \sum _{k=0} ^\infty 
    {\Gamma (k+{3\over 2}) \over \Gamma ({3\over 2})\,  k!} 
    {(1-t^2) ^k \over 2k+n +2}
\cr
&=
-{t \over 2 +n} \, \ln {1-t^2 \over 4} \,
    F\big ({3\over 2}, {n \over 2} +1; {n \over 2} +2; 1-t^2\big ).
\cr}
\eqno (4.31)
$$
The functions $a^{(n)} _{\rm reg} (t)$ admit a Taylor series expansion in
powers of $1-t^2$, given as follows
$$
\eqalign{
a^{(n)} _{{\rm reg}} (t)
& =
2t \sum _{k=0} ^\infty {a_k \over 2k+n +2} (1-t^2)^k 
\cr
a_k 
& =
-1 +{\Gamma (k+{3 \over 2}) \over \Gamma ({3\over 2}) \, k!}
\Bigl \{ {1 \over 2k+n+2} - {1 \over \pi} \sum _{m=1} ^k 
    {\Gamma (m+\half) \Gamma (k-m +{3\over 2}) 
    \over 
    \Gamma (k-m+1) \cdot m \cdot m!} \Bigr \}.
\cr}
\eqno (4.32)
$$
It is understood that in the expression for $a_0$, the last sum term 
does not contribute.
Using this expansion, one may now evaluate $I^{(\ell)}$, given by (4.8), and
from there the reduced amplitude $B(x_i)$ using (4.2), and the full amplitude
$A(x_i)$ using (3.10). These calculations are elementary and we shall not carry
then out here. Suffice it to note that the leading logarithmic behavior
identified in (4.25) and (4.26) is precisely recovered from the term
proportional to $\ln (1+t)$ of $I(s,t)$ in (4.29) and (4.31).

\bigbreak
\centerline{{\bf 5. Conclusions and Outlook}}

\bigskip
\noindent
In this paper we have studied the contribution of the gauge bosom exchange
diagram in $AdS_{d+1}$ to the boundary correlator $\left\langle \O_\Delta (x_1) 
\O_{\Delta'} (x_2)
     \O^*_\Delta (x_3) \O^*_{\Delta'} (x_4) \right\rangle _{\rm gauge}$
of charged scalar composite operators of arbitrary integer dimension. The
calculation was feasible because we used a simple new form of the
gauge boson propagator derived in \S 2 by a covariant method in which
physical effects were separated from gauge artifacts. 
Although the results hold for general even boundary dimension $d$, we shall
discuss them  for the case of primary interest, namely $d=4$, where the
boundary conformal theory is $\N=4$ super--Yang-Mills theory.

\medskip

The gauge boson exchange process, for the gauge group $SU(4)$ of the bulk
$\AdS_5 \times S_5$ supergravity theory, is a fundamental sub-process in many
4-point correlators of interest, although it does not give the
complete amplitude for any correlator. So a full analysis of 4-point
functions must wait until we learn how to compute exchange diagrams
with massive scalars and vectors and as well as the graviton and massive
tensors. Explicit bulk-to-bulk propagators must be found for most of
these cases. Once they are found we believe that the calculational methods
developed here can be applied to calculate the bulk integrals. Explicit
cubic and quartic couplings of the supergravity theory will also be
required.

\medskip

We have shown that the gauge boson exchange diagram contains logarithmic
singularities of the type [7] previously found for 4-point contact
interactions. The logarithmic singularities are the leading contribution
in the $s$-channel limit, $|x_{12}| \to 0$, but they occur at non-leading
level in the $t$-channel, $|x_{13}| \to 0$. The leading singularities in
the $t$-channel are powers which correspond to a simple double OPE type 4-point
function [9] in which the conserved $SU(4)$ flavor current of the 
boundary theory and its descendents appears in the intermediate state.
However if logarithms do not cancel with those in other diagrams, 
complete 4-point correlators will be more complicated than the double 
OPE form in (1.1) (rewritten for the $t$-channel).

\bigbreak
\centerline{{\bf Acknowledgments}} 
\frenchspacing

\bigskip
\noindent
It is a pleasure to acknowledge helpful conversations with Samir Mathur and 
Leonardo Rastelli. We also wish to thank the Aspen Center for Physics, where 
most of this work was carried out.

\bigbreak

\centerline{{\bf References}}

\item{[1]} J.M.  Maldacena, ``The Large N Limit of Superconformal Field Theories
      and Supergravity, hep-th/9711200.

\item{[2]} S. Gubser, I.R. Klebanov and A.M. Polyakov, 
``Gauge Theory Correlators from Non-Critical String Theory",
Phys. Lett {\bf B428} (1998) 105; hep-th/9802109.

\item{[3]} E. Witten, ``Anti De Sitter Space And Holography'',
hep-th/9802150.

\item{[4]} W. Mueck and K.S. Viswanathan, ``Conformal Field Theory Correlators 
  from Classical Scalar Field Theory on $AdS_{d+1}$ ", 
  Phys. Rev. {\bf D58} (1998) 41901; hep-th/9804035.

\item{[5]} D.Z. Freedman, A. Matusis, S. D. Mathur, and L. Rastelli, as
discussed in Freedman's conference lecture at Strings '98, 
 http://www.itp.ucsb.edu/online/strings98/.

\item{[6]} H. Liu and A.A.  Tseytlin, ``On Four-point Functions in the
CFT/AdS Correspondence''; hep-th/9807097.

\item{[7]} D.Z. Freedman, A. Matusis, S. D. Mathur, and L. Rastelli,
 ``Comments on 4-point functions in the CFT/AdS correspondence'',
  hep-th/9808006.

\item{[8]}  J.H. Brodie and M. Gutperle, ``String Corrections to four 
point functions in the AdS/CFT correspondence", hep-th/9808067.

\item{[9]} S. Ferrara, conference lecture Strings '98, 
    {http://www.itp.ucsb.edu/online/strings98/}; S. Ferrara, R. Gatto, 
 A.F. Grillo, and G. Parisi, Nucl. Phys. {\bf B49} (1972) 77; Nuovo Cimento
 {\bf 26} (1975) 226

\item{[10]} S. Lee, S. Minwalla, M. Rangamani, and N. Seiberg,
  ``Three-Point Functions of Chiral Operators in D=4, $\N=4$ SYM at Large N'';
   hep-th/9806074.

\item{[11]} E. D'Hoker, D.Z. Freedman, and W. Skiba, ``Field Theory Tests for 
Correlators in the AdS/CFT Correspondence''; hep-th/9807098.

\item{[12]} H.J. Kim, L.J. Romans and P. van Nieuwenhuizen, ``Mass spectrum of 
the chiral ten-dimensional $N=2$ supergravity on $S^5$", Phys. Rev. {\bf D32} 
(1985) 389.

\item{[13]} C. Fronsdal, Phys. Rev. {\bf D10} (1974) 589;
C.P. Burgess  and C. A. Lutken, Propagators and Effective Potentials in
Anti-de Sitter Space'', Nucl. Phys. {\bf B272} (1986) 661;
T.  Inami and H. Ooguri, ``One Loop Effective Potential in Anti-de Sitter 
Space'', Prog. Theo. Phys. {\bf 73} (1985) 1051;
C.J. Burges, D.Z. Freedman, S. Davis, and G.W. Gibbons, ``Supersymmetry
in Anti-de Sitter Space'', Ann. Phys. {\bf 167} (1986) 285.

\item{[14]} B. Allen and T. Jacobson, ``Vector Two Point Functions in
Maximally Symmetric Spaces'', Commun. Math. Phys. {\bf 103} (1986) 669. 

\item{[15]} D.Z. Freedman, A. Matusis, S. D. Mathur, and L. Rastelli,
``Correlation functions in the CFT(d)/AdS(d+1) correspondence'';
  hep-th/9804058.

\bye